\documentclass[twocolumn,showpacs,prl,aps]{revtex4-1}

\usepackage{dcolumn}
\usepackage{amssymb}
\usepackage{bm}

\draft

\begin{document}

\title{Some comments to the quantum fluctuation theorems}

\author{Yu. E. Kuzovlev}
\email{kuzovlev@fti.dn.ua} \affiliation{Donetsk
Physics and Technology Institute, 83114 Donetsk, Ukraine}


\begin{abstract}
It is demonstrated that today's quantum fluctuation theorems %
are component part of old quantum fluctuation-dissipation relations %
[Sov.Phys.-JETP  45, 125 (1977)], and typical %
misunderstandings in this area are pointed out.
\end{abstract}


\maketitle

{\bf 1}.\,\, This communication is devoted to %
those general statistical properties of externally %
driven quantum systems what are due to time reversibility %
of their underlying (Hamiltonian) microdynamics. %
For the first time a complete list of such properties %
was suggested in item 5 in \cite{bk1}. Later %
\cite{bk2,bk3} they wee named ``generalized %
fluctuation-dissipation relations'' (FDR) %
since relate dissipation and irreversibility %
in (linear and non-linear) response functions %
to fluctuations (see also %
\cite{fdrcm,fdrn} and references therein). %

Here, I would like to comment %
last decade activity in this field, %
concerning the quantum ``fluctuation theorems'' %
(FT) \cite{ehm,ag,cht}, %
and focusing on three its aspects as follow. %
Firstly, that some authors surprisingly do not understand %
existence of two qualitatively different types %
of the external Hamiltonian driving. %
Secondly, that some recently presented relations %
in fact are paricular cases of the old FDR. %
Thirdly, that the only real problem in this %
field is establishing of physically meaningful %
rules of ordering and symmetrization of %
quantum operators and super-operators.

{\bf 2}.\,\, %
As in \cite{fdrn}, let us start from %
separation of two types of external driving. %
One is such that small changes of %
a driving ``force''\, $\,x=x(t)\,$\, cause %
only small changes in any degree of freedom %
or part of our (now quantum) system. %
Another type is such that even small change in %
\,$\,x\,$\, can cause arbitrary strong changes %
in some degrees of freedom and thus strong energy %
exchange between them and the rest of the system. %
For example, if system represents quasi-free %
charge carrier with charge $\,q\,$ %
in dielectric fluid (or %
crystal) contained in volume with diameter $\,L\,$, %
and $\,x\,$ is electric field, then equilibrium %
state at any $\,x=\,$const$\,\neq 0\,$ %
(let arbitrary small) realizes only %
when the carrier achieves boundary of the container %
while the force lowers the carrier's potential energy %
by $\,w\sim |qx|L\,$ thus producing work %
$\,w\,$ dissipated by the fluid (crystal). %

Clearly, on one hand, free energy of this final %
state significantly differs from starting value %
(at $\,x=0\,$). On the other hand, %
if we are interested in the carrier's %
transport in itself, then both the final equilibrium %
state and its free energy $\,F(x)\,$ %
are far from of our interest. In this case, %
instead, we have to consider thermodynamic limit %
$\,L\rightarrow \infty\,$ (or not too long time %
intervals only) and thus non-equilibrium steady %
(or quasi-stationary) state. %

Hence, the corresponding FDR characterize %
essentially non-equilibrium transient process, %
either steady or finishing at essentially new state %
with different free energy. %
But in no way %
``the restricted case where there %
is no change in free energy'' (citation %
from \cite{ag}, p.230404-2, right column) ! %
Such curious view \cite{ehm,ag} %
says about incomprehension of  %
true contents of FDR (and hence FT) %
and headless reading of \cite{bk1}. %

We see that under the second type of driving %
a part of the whole system (charge carrier %
in the above %
example) behaves as open system (in between %
of driving source and other parts may be playing %
role of  thermostats). Therefore %
below for brevity let us call this case ``open'' %
while the first one ``closed''. In ``open'' case %
a natural measure of violation of equilibrium by %
external force $\,x(t)\,$ is its value itself %
while in ``closed'' case a value of its time %
derivative, $\,dx(t)/dt\,$. This difference %
can be underlined by different representations %
of FDR \cite{bk3} (though, of course, %
it wipes out under sufficiently %
high-frequency oscillating $\,x(t)\,$).

{\bf 3}.\,\, %
In \cite{bk1} an unified consideration %
of both the ``closed''  and ``open'' cases was %
suggested, using division of full system's %
Hamiltonian, $\,H(x)\,$, as follows:
\begin{equation}
\begin{array}{c}
H(x)=H_0-h(x)\,\,,\,\,\,\,\, \,\, %
H_0(x)=H(x_0)\,\,,\,\\ %
h(x)=H(x_0)-H(x)\,\, %
\label{h}
\end{array}
\end{equation}
(page 127, left column in \cite{bk1}). %
In the open case, obviously, it is %
reasonable to put on $\,x_0=0\,$ and %
besides write the ``perturbation Hamiltonian'' %
$\,h(x)\,$ in the form $\,h(x)=xQ\,$ %
(although the latter is not necessary). %
Main relation of \cite{bk1}, expressed by %
formulae (17),(19),(21) and (23) there, is %
\begin{eqnarray}
\langle\, A_1(t_1)\dots A_n(t_n)\, %
e^{-\beta H_0(t)}e^{\,\beta H_0(0)}\, \rangle %
_{x(\tau)}\,=\,\nonumber\\ %
=\, %
\langle\, \overline{A}_n(t-t_n)\dots %
\overline{A}_1(t-t_1)\, %
\, \rangle_{\epsilon x(t-\tau)}\,\,\,,
\label{mr}
\end{eqnarray}
where\, $\,A_j\,$ are arbitrary operators,\, %
the over-line means their transposition,\, %
presence of time arguments in $\,A_j(t_j)\,$ and %
$\,H_0(t)\,$ means that operators are treated in %
the Heisenberg picture,\, %
$\,H_0(0)=H_0\,$,\, %
$\,\epsilon\pm 1\,$ is %
paritity (or parities) of the driving force %
(or forces) $\,x(t)\,$  in respect to time reversal,\, %
and angle brackets denote average over canonical %
distribution of initial conditions under given %
(arbitrary) trajectory of the force:
\begin{equation}
\begin{array}{c}
\langle\dots\rangle_{x(\tau)} = %
\texttt{Tr\,} \dots \rho_0\,\,,\,\,\,\,\, %
\rho_0 = q^{-1}\exp{(-\beta H_0)}\,\,, %
\label{r0}
\end{array}
\end{equation}
with\, $\,q\,$ being nozmalazing factor. %
At that, if $\,H_0\,$ includes magnetic field %
(or other time-odd parameter) then on the %
right in (\ref{mr}) it must be inverted %
(\cite{bk1}, p.129, left column). %
If $\,A_j\,$ are Hermitian and possess %
definite parities, %
$\,\overline{A}_j=\epsilon_j A_j\,$, %
then (\ref{mr}) turns to formula (24) %
from \cite{bk1},
\begin{eqnarray}
\langle\, A_1(t_1)\dots A_n(t_n)\, %
e^{-\beta H_0(t)}e^{\,\beta H_0(0)}\, \rangle %
_{x(\tau)}\,=\, \nonumber\\ %
=\, %
\epsilon_1\dots\epsilon_n\, %
\langle\, A_n(t-t_n)\dots A_1(t-t_1)\, %
\, \rangle_{\epsilon x(t-\tau)}\,
\label{mr1}
\end{eqnarray}
Formulas (\ref{mr}) or (\ref{mr1}) represent %
complete lists of symmetry relations produced %
by the microscopic time reversibility and %
therefore can be termed ``generating quantum FDR''. %

{\bf 4}.\,\, %
Of course, these FDR are more comfortable  %
in the ``open'' case than in `closed'' one (see item %
2 above). Therefore, if we are ready to restrict %
our consideration by closed case (and thus closed %
systems) only, then it is suitable to replace %
$\,H_0\,$ in (\ref{mr})-(\ref{mr1}) by $\,H(x)\,$. %
At that, the only additional difference from %
derivation of (\ref{mr}) and (\ref{mr1}) on p.129 %
in \cite{bk1} is that normalizing factor %
$\,q\,$ becomes $\,x\,$-dependent. Repetition of %
the derivation yields
\begin{eqnarray}
\langle\, A_1(t_1)\dots A_n(t_n)\, %
e^{-\beta H(t,x(t))}e^{\,\beta H(0,x(0))}\, \rangle %
_{x(\tau)}\,=\,\nonumber\\ %
=\, %
\langle\, \overline{A}_n(t-t_n)\dots %
\overline{A}_1(t-t_1)\, %
\, \rangle_{\epsilon x(t-\tau)} %
\,\times\, \label{nr}\\
\times\,\, e^{\,\beta [F(x(0))-F(x(t))]} %
\,\,\,, \nonumber
\end{eqnarray}
where\, $\,H(t,x(t))\,$ means $\,H(x(t))\,$ %
taken in the Heisenberg oicture, %
so that $\,H(0,x)=H(x)\,$,\, %
\begin{equation}
\begin{array}{c}
\langle\,\dots\,\rangle_{x(\tau)} = %
\texttt{Tr\,} \dots\, \rho(x(0))\,\,\,, \label{r}\\ %
\rho(x) = q^{-1}(x)\,\exp{(-\beta H(x))}\,\,, %
\end{array}
\end{equation}
and\, $\,F(x)=-\beta^{-1} \ln{\,q(x)}\,$\,.
If all $\,A_j\,$ are Hermotian with definite %
parity, then
\begin{eqnarray}
\langle\, A_1(t_1)\dots A_n(t_n)\, %
e^{-\beta H(t,x(t))}e^{\,\beta H(0,x(0))}\, \rangle %
_{x(\tau)}\,=\,\nonumber\\ %
=\, %
\epsilon_1\dots \epsilon_n\, %
\langle\, A_n(t-t_n)\dots A_1(t-t_1)\, %
\, \rangle_{\epsilon x(t-\tau)}\,\times\, %
\label{nr1}\\
\times\,\, e^{\,\beta [F(x(0))-F(x(t))]}  %
\,\, \nonumber
\end{eqnarray}

Notice that FDR (\ref{nr}) and \ref{nr1}) can be %
obtained also directly from FDR  %
(\ref{mr}) and \ref{mr1}), respectively, %
as their particular (``closed'') case, in full %
analogy with transition between two classical %
FT in \cite{fdrn}. %
It is sufficient, at %
any fixed $\,t\,$ and $\,x(t)\,$ %
to choose in (\ref{h})\, %
$\,x_0=x(t)\,$\, and then %
apply identity
\begin{eqnarray}
e^{-\beta H_0(t)} e^{\,\beta H_0} %
\rho_0= \frac {q(x(0))}{q(x(t))}\, %
e^{-\beta H(t,x(t))} e^{\,\beta H(x(0))}\, %
\rho(x(0))\,\,, \nonumber
\end{eqnarray}
where the above designations also are used.

{\bf 5}.\,\, %
In general, any rearrangement of operators %
$\,A_j(t_j)\,$ in the relations (\ref{mr}), (\ref{mr1}), %
(\ref{nr}) and (\ref{nr1}) produces, - %
in contrast to the classical theory, - %
not identical but new relation. Therefore, any %
a priory prescribed re-ordering or symmetrization of %
these operators, in respect to their time %
arguments or indices, abolishes a part of information %
contained in initial relations and thus lowers %
generality of result as compared with them. %
Consequently, any grouping of particular FDR %
(\ref{mr}), (\ref{mr1}), (\ref{nr}) or (\ref{nr1}) %
into some generating relation, e.g. for some %
characteristic function or functional, leads to %
loss of information and instead of heightening %
generality (as it would be in classical theory) %
in fact lowers it! %

For example, let us make in (\ref{nr1}) %
redesignation $\,A_j\Rightarrow A_{k_j}\,$, %
multiply both sides by %
$\,\prod_{j=1}^n u_{k_j}(t_j)\,$, where %
$\,u_{k}(t)\,$ are arbitrary (complex) test %
functions, take sum over all $\,k_j\,$, %
integrate over all $\,t_j\,$, divide by $\,n!\,$ %
and sum over all non-negative $\,n\,$. These %
operations can be easy performed mentally, ``without %
paper and pencil'', producing %
generating FDR %
\begin{eqnarray}
\langle\, \exp{[\int_0^t u_k(\tau) %
A_k(\tau)\,d\tau ]}\, %
e^{-\beta H(t,x(t))}e^{\,\beta H(0,x(0))}\, \rangle %
_{x(\tau)}\,=\, \nonumber\\ %
\,=\langle\, \exp{[\,\epsilon_k %
\int_0^t u_k(t-\tau) %
A_k(\tau)\,d\tau ]}\, %
\, \rangle_{\epsilon x(t-\tau)} %
\,\times \,\nonumber\\
\times\,\,e^{\beta [F(x(0))-F(x(t))]}\, %
\,\,\,\,\,\, \label{nr2}
\end{eqnarray}
(with summation over repeated index $\,k\,$). %
But what it really generates (after functional %
differentiations by $\,u_k(\tau)\,$) ? %
Clearly, relations for fully symmetrized %
quantum %
statistical moments only, thus losing an unknown %
amount of information in comparison %
with (\ref{nr1}) (e.g. information from various %
commutators of $\,A_k(t)\,$).

Hence, the above relations for arbitrary
asymmetric non-ordered quantum moments are most general %
form of FDR. Just by this reason authors of \cite{bk1} %
confined themselves by non-symmetrized relations  %
(but not because they ``were not able'' to perform %
the mentioned trivial operations, as assumed in %
\cite{cht}). Therefore it looks strangely when authors %
of \cite{ag,cht} characterize %
relations like (\ref{nr2}) (formula (12) %
from \cite{ag} or (55) from \cite{cht}) as ``universal'' %
and ``most general''. %

In reality, relations (\ref{nr2}) can not generate even %
the Efremov's quadratic FDT \cite{efr} or complete %
set of independent four-index %
quantum FDR \cite{strb}. %
If one wants to avoid loss of these %
results and simultaneously use fully %
symmetrized moments only, then the latter  %
should be defined like it was suggested %
in Sec.3 in \cite{bk3} (or see formula (2) %
in \cite{fdrcm}).

{\bf 6}.\,\, %
Actual achievement of quantum theory in the field %
of FDR after \cite{bk1} is not formulae like %
(\ref{nr2}) but realization of those %
circumstance that measuring of energy difference %
between two system's states requires %
two independent measurements of energy, %
therefore, operator $\,H(t,x(t))-H(0,x(0))\,$,  %
or $\,H_0(t)-H_0(0)\,$, %
is not quantum %
observable representing %
the energy difference \cite{cht}. %

In other words, the product of two %
exponentials in (\ref{mr}) and (\ref{mr1}), %
or (\ref{nr}) and (\ref{nr1}), is true %
quantum equivalent of classical exponentials %
$\,\exp{(-\beta E)}\,$ or %
$\,\exp{(-\beta \mathcal{E})}\,$, respectively, %
with $\,E= H_0(t)-H_0(0)\,$ being system's internal %
energy change in the open case and %
$\,\mathcal{E}=H(t,x(t))-H(0,x(0))\,$ %
total energy change in the closed case. %

Hence, characteristic function of energy change %
also must be defined by means of %
pair of exponentials \cite{ehm,cht}, for instance, %
\begin{equation}
G(u;x(\tau))\,=\,%
\langle e^{\,uE}\rangle_{x(\tau)}\,\equiv\, %
\texttt{Tr\,}\, %
e^{\,uH_0(t)}e^{-uH_0(0)}\rho_0\,\, \label{def}
\end{equation}
Then, choosing in (\ref{mr}) %
$\,n=2\,$, $\,A_1=\exp{(-uH_0)}\,$, %
$\,A_2=\exp{(uH_0)}\,$, $\,t_1=0\,$ %
and $\,t_2=t\,$, %
one obtains characteristic function %
representation of the ``quantum work %
FT'' \cite{ehm,cht}:
\begin{equation}
G(u-\beta;x(\tau))\,= %
\,G(-u;\epsilon x(t-\tau)) \,\, \label{qw}
\end{equation}
Quite similarly from (\ref{nr}) arises %
FT for the\, $\,\mathcal{E}\,$\,. %

{\bf 7}.\,\, %
The same can be said about differences $\,Q(t_f)-Q(t_i)\,$ %
of any other quantum observable %
(e.g. space coordinate of charge carrier in example from %
item 2 above). %
However, since generally $\,Q\,$ does not %
commute with $\,H_0\,$ and $\,H(x)\,$, %
we can not merely replace $\,H_0\,$ in %
(\ref{def}) by $\,Q\,$, %
but have to define a reasonable %
rule for ordering or/and symmetrization %
of operator products. %

Among many formally acceptable rules %
there is one prompted by the %
correspondence principle. %
It states that %
characteristic functional of %
any variable (or set of variables) $\,A\,$ %
can be represented (see e.g. \cite{fdrcm} %
and references therein) by
\begin{equation}
\begin{array}{c}
\langle\, %
\exp{\,\int_0^t u(\tau)A(\tau)\, d\tau } %
\,\rangle_{x(\tau)}\,\,%
=\, \texttt{Tr\,} \rho\,\,\,, \label{cf}
\end{array}
\end{equation}
where\, $\,\rho\,$ %
is solution to equation %
\begin{equation}
\begin{array}{c}
\dot{\rho}\,=\,[H(x(t)),\rho ]/i\hbar %
\,+\,u(t)\,A\circ\rho \,\,\,,
\label{ce}
\end{array}
\end{equation}
with\, $\,\circ\,$\, denoting %
the symmetrized, or Jordan, product, %
$\,A\circ B\equiv (AB+BA)/2\,$. %
The corresponding rule %
is chronological time ordering of %
Jordan products: %
%
\begin{equation}
\begin{array}{c}
\langle\, %
\exp{\,\int_0^t u(\tau)A(\tau)\, d\tau } %
\,\rangle_{x(\tau)}\,\,=\, %
\, \label{pf}\\%
=\, \texttt{Tr\,}\,\overleftarrow{\exp} %
[\frac 12 \int_0^t %
u(t^{\prime})A(t^{\prime})dt^{\prime} %
]\rho _0\,\, %
\overrightarrow{\exp}[\frac 12 \int_0^t %
u(t^{\prime})A(t^{\prime}) %
dt^{\prime}]\,
\end{array}
\end{equation}
with $\,\overleftarrow{\exp}$ and %
$\overrightarrow{\exp}$ being %
chronological and anti-chronological %
exponents. %
According to this rule, %
\begin{equation}
\langle A_1(t_1)\dots %
A_n(t_n)\rangle_{x(\tau)} = %
\texttt{Tr\,} \mathcal{T}\,\left[ %
\prod_{j=1}^n  A_j(t_j)\circ %
\right]\rho _0\,\,\,, \label{m}
\end{equation}
where $\,\mathcal{T}\,$ means chronological %
ordering of the Jordan products, and  %
\begin{eqnarray}
\langle e^{u[Q(t_f)-Q(t_i)]}\rangle_{x(\tau)}\,= %
\,\,\, \label{d}\\
=\texttt{Tr\,}\, e^{\,uQ(t_f)/2} %
e^{-uQ(t_i)/2}\,\rho_0\, %
e^{-uQ(t_i)/2} e^{\,uQ(t_f)/2}\, %
\nonumber
\end{eqnarray}
under\, $\,t_f>t_i\geq 0\,$. %
Clearly, this rule relates any %
finite-time difference to two separate %
measurements. In particular, %
at $\,Q=H_0\,$ (and $\,t_i=0\,$, %
$\,t_f=t\,$) expression (\ref{d}) %
coincides with (\ref{def}).

{\bf 8}.\,\, %
Derivation of FDR under just defined %
ordering-symmetrization rule copies %
derivation of (\ref{mr}), %
(\ref{mr1}) in \cite{bk1}. %
For the open case, the result is
\begin{eqnarray}
\langle\,e^{-\beta H_0(t)}\, %
\exp{[\int_0^t u_k(\tau)A_k(\tau)\, d\tau ]} %
\,e^{\,\beta H_0(0)} %
\,\rangle_{x(\tau)}\,\,=\, %
\nonumber \\%
=\, \langle\, %
\exp{[\int_0^t \epsilon_k %
u_k(t-\tau)A_k(\tau)\, d\tau ]} %
\,\rangle_{\epsilon x(t-\tau)}\, %
\,\,\,\,\, \label{fdr}
\end{eqnarray}
For transition to the closed case one has only to %
replace $\,H_0(t)\,$ by $\,H(t,x(t))\,$and add %
to the right side multiplier %
$\,\exp\,\beta[F(x(0))-F(x(t))]\,$. %
It is easy to see that ``quantum work FT'' %
under this rule in fact coincides with (9)-(10).

In many applications of FDR one would like to connect %
the pair\, $\,e^{-\beta H_0(t)}\,\dots \,e^{\,\beta H_0(0)}\,$, - %
i.e. energy changes, - to some of variables $\,A_k(t)\,$ %
under attention. %
In particular, when considering %
connections between fluctuations and non-linear %
responses at given ordering-symmetrization rule. %
Careful (though not complete yet) %
analysis of FDR (\ref{fdr}) %
from such viewpoint was undertaken in \cite{fdrcm}.

{\bf 9}.\,\, %
In conclusion I should notice that the %
above considered relations can be extended %
to ``thermic perturbations'' \cite{bk3} %
(i.e. non-equilibrium perturbations of initial %
state of the system, in addition to its %
Hamiltonian driving), thus expanding variety %
of FDR's applications to %
non-equilibrium processes in open systems (see %
\cite{bk2,bk3} and references from \cite{fdrn}). %



\begin{thebibliography}{10}

\bibitem{bk1}
G.N.Bochkov and Yu.E.Kuzovlev, %
``General theory of thermal
fluctuations in nonlinear systems'',\, %
Sov.Phys.-JETP {\bf 45}, 125 (1977);\, %
\url{http://www.jetp.ac.ru/cgi-bin/dn/e_045_01_0125.pdf}


\bibitem{bk2}
G.N.Bochkov and Yu.E.Kuzovlev,\, %
``Fluctuation-dissipation relations
for non-equilibrium processes in open systems'',\, %
Sov.Phys.-JETP {\bf 49}, 543 (1979);\, %
\url{ttp://www.jetp.ac.ru/cgi-bin/dn/e_049_03_0543.pdf}

\bibitem{bk3}
G.N.Bochkov and Yu.E.Kuzovlev,\, %
``Non-linear %
fluctuation-dissipation relations and stochastic models in %
non-equilibrium thermodynamics. %
I. Generalized fluctuation-dissipation theorem'',\, %
Physica {\bf A 106},\, 443\, (1981).


\bibitem{fdrcm}
Yu.E.Kuzovlev, \,``Fluctuation-dissipation relations for %
continuos quantum measurements'',\, %
arXiv:\, cond-mat/0501630\,.

\bibitem{fdrn}
Yu.E.Kuzovlev, \,``Short remarks on the so-called fluctuation %
theorems and related statements'',\, %
arXiv:\, 1106.0589\,.


\bibitem{ehm}
M.\,Esposito, U.\,Harbola, and S.\,Mukamel,\, %
``Nonequilibrium fluctuations, fluctuation %
theorems, and counting statistics in quantum systems'',\, %
Rev. Mod. Phys. {\bf 81}, 1665 (2009).

\bibitem{ag}
D.\,Andrieux and P.\,Gaspard, ''Quantum work relations and %
response theory'',\. Phys. Rev. Lett. {\bf 100}, 230404 (2008).

\bibitem{cht}
M.\,Campisi, P.\,Ha?nggi, and P.\,Talkner,\, %
``Colloquium: Quantum fluctuation relations: foundations %
and applications'',\, Rev. Mod. Phys. {\bf 83}, 771 (2011).


\bibitem{efr}
G.F.Efremov,\, ``A fluctuation dissipation theorem for nonlinear %
media'',\, Sov. Phys. JETP {\bf 28}, 1232 (1969).


\bibitem{strb}
R.L.Stratonovich. Nonlinear nonequilibrium \\thermodynamics.\, %
Springer Series in %
Synergetics, Vol.\,59.\, Springer-Verlag, Berlin.




\end{thebibliography}



\end{document}